\documentclass[12pt]{article}
 \newcommand {\be} {\begin{equation}}
\newcommand {\bea} {\begin{eqnarray} \nonumber }
\newcommand {\ee} {\end{equation}}
\newcommand {\eea} {\end{eqnarray}}

 \newcommand {\Si} {\Sigma}
\newcommand {\om} {\omega}
\newcommand {\de} {\delta}
\newcommand {\g} {\gamma}

\newcommand {\la} {\lambda}
\newcommand {\La} {\Lambda}
 \newcommand {\al} {\alpha}
 
 \newcommand {\N} {{\cal N}}
\newcommand {\R} {{\cal R}}
\newcommand {\D} {{\cal D}}
\newcommand {\PP} {{\cal P}}

\newcommand {\ba} {\overline}
\newcommand {\lan} {\langle}
\newcommand {\ran} {\rangle}

\newcommand {\Tr} {\mbox{Tr}}
\newcommand {\for} {\ \ \ \mbox{for}\ \ \ }

\def \form#1 {eq. (\ref{#1}) }

\begin{document}
	\title{On the most compact regular lattice in large dimensions: A statistical mechanical 
	approach}
	\author{Giorgio Parisi\\Dipartimento di Fisica, Sezione INFN, SMC of INFM-CNR,\\
Universit\`a di Roma ``La Sapienza'',\\
Piazzale Aldo Moro 2,
I-00185 Rome (Italy)}
	\maketitle
	\abstract{In this paper I will approach the computation of the maximum density of regular lattices in large dimensions using a statistical mechanics approach.
	The starting point will be some theorems of Roger, which are virtually unknown in the community of physicists. Using his approach one can see that there are many similarities 
	(and 
	differences) with the problem of computing the entropy of a liquid of perfect spheres. The relation between the two problems is investigated in details. Some conjectures are
	presented, that need further investigation in order to check their consistency.}

\section{Introduction}
The aim of this paper is to study a well known and celebrated problem, i.e. which is the 
maximum density of hard spheres when they are packed on a regular lattice.  In two and 
three dimensions the solution is well known; the lattices with maximal packing density are 
the hexagonal and the ffc lattices respectively.  

In generic dimensions the result for the maximal packing density in not known \cite{CON}: 
an lower bound on the maximal packing density has been established by Minkowski 
\cite{MINK} and it has only marginally improved by later studies.

In this paper, after a brief review of the established mathematical results, 
we will show how the usual techniques of statistical mechanics may be used 
in this problem.  We will not be able to find the maximal packing lattice 
density in high dimensions, but we believe that our work may be an useful 
step in this direction, especially for pointing the connections between this 
problem and the statistical mechanics disordered systems, like spin glasses and structural glasses.  As we shall see later, 
there are very interesting relations among this problem and the 
thermodynamic properties of an hard sphere liquid.

In section II we shall introduce the basic concepts and definitions which will be used later, in particular we will introduce the Roger's measure over the space of all possible
lattices.  In section III we will present the main mathematical results, due to Roger \cite{ROGER}, in particular we will show how to compute the moments of a function of the
lattice using Roger's measure.  In section IV we perform some simple computations in order to become more familiar with Roger's results.  In section V we show how the whole
approach can be simplified if we introduce connected correlations functions.  In section VI we present a crucial technical conjecture that is needed to make further progresses.
Finally in the last section we start a comparison between the the packing problem and the hard sphere liquid; we present a possible conjecture for the relation between these two
problems in the the infinite dimensional limit.  Further work is needed to find out if the conjecture is consistent.  Finally we present our conclusions and some comments on
possible developments.  An appendix is devoted to the computation of some integrals in the infinite dimensional limit.

\section{Some definitions}
Let us start by defining the problem and by establishing our notations. The first 
choice is how to parametrize the regular lattices having cells of unit volume.

This can be done by considering the set of unimodular real square matrices $D \times D$ 
(i.e. those matrices with determinant equal to 1).  
To each matrix $\La$ of this set we can associate a regular lattice, that is given by all 
the points of the form
\be
x_i=\sum_k \La_{i,k} n_k\equiv(\La n)_i,
\ee
where the $n_k$ take all the possible integer values (both positive and negative).  This 
correspondence is not one to one: the same lattice can be obtained by infinitely many 
different matrices.

The condition of unimodularity (i.e. $\det(\La)=1$) implies that the volume of the fundamental cell
of the lattice is also equal to one. 

If we consider non overlapping spheres centered on the points of the lattices, the maximum allowed
{\sl diameter }($\R(\La)$) is given by
\be
\R(\La)^2 =\min_n |\La n|^2 = \min_n \sum_{i,k} n_i A_{i,k} n_k,\label{MINI}
\ee
where the positive matrix $A$ is given by
\be
A_{i,k} = \sum_j \La_{i,j} \La_{k,j}
\ee
and the minimum is done over all the possible choice of the integer 
 $D$-dimensional vector $n$, with the exclusion of the
origin. Indeed the quantity defined in eq. (\ref{MINI}) is the minimum distance of a point of 
the lattice from the origin.

Our aim is to compute 
\be
\R_M= \max_\La \R(\La).
\ee

We remark that the computation of the minimum in equation (\ref{MINI}) for generic $\La$ 
or $A$ is an NP-hard problem for large dimensions $D$.  Of course there are some matrices (e.g. 
diagonal ones) for which the computation is quite simple. On the other end, in cases 
where the off diagonal elements are large, the computation may becomes quite complex.
This fact may suggest that some of the 
techniques used in spin glasses (where one  studies the statistical mechanics of some NP 
complete problems \cite{BOOK}) may be relevant also in this case.

In order to analyze better this problem we introduce the notation
\be
\lan f(|x|)\ran_\La \equiv \sum_n f(|x(n)|) \equiv \sum_n f(|\La n|),
\ee
where the sum is done over $Z^D$, origin excluded.

The quantity $\R(\La)$ can computed as
\bea
\R(\La)^2 = -\lim_{\beta \to \infty} \ln( F(\beta)_{\La})\\
F(\beta)_{\La}={\lan\exp ( -\beta |x|^2) \ran _\La \over \beta}.
\eea
In other words $|x|^2$ is the Hamiltonian, 
\be
Z(\beta)_\La \equiv \lan\exp ( -\beta |x|^2) 
\ran_\La \label{PART}
\ee
is the partition function and $\R(\La)^2$ is the ground state energy.  

The 
typical problem in spin glasses would be the computation of the average value (over $\La$) 
of the free energy associated to the previous defined partition function.
This is not the problem we face here.  In this language the maximum packing density problem 
is a minimax problem, i.e. it consists in finding the matrix $\La$ (or equivalently $A$) such 
the ground state energy is as large as possible. This problem has been recently addressed for spin glasses in ref. \cite{PR}.

We can formulate the same problem in an alternative way if we consider the quantity
\be
K(R)_\La= \lan \theta(R-|x|)\ran_\La, \label{SR}
\ee
i.e.  the number of points, excluded the origin, inside a sphere of 
radius $R$.
It is obvious that
\bea
K(R)_\La=0 \for R<\R(\La),\\
K(R)_\La>0 \for R>\R(\La).
\eea
Indeed the {\sl radius} of the largest sphere that is centered in one point does not intersect the 
other points is the {\sl diameter} of the largest spheres, that do not overlap, if they are 
centered on the points of the lattice.

If at given $\La$ we were able to compute $\lan \cdot \ran_\La$, we would immediately 
obtain the wanted result after a maximisation over $\La$ of the appropriate quantity.  As 
we have already remarked the computation of these expectation values for a given lattice is a nasty 
problem 
that cannot simply solved analytically.  As 
usual in the case of random problems we introduce an ensemble of problems and we try to 
compute the average over this ensemble.  This is usually a goal that we can reach 
analytically.

In order to formulate the statistical problem in a more precise way we must firstly 
a measure $d\mu(\La)$ on the space of all possible lattices;  we will specify later the 
form of this measure.  We define
\be
\ba{ F(\La)} = \int d\mu(\La) F(\La).
\ee

If the measure $d\mu(\La)$ does not vanish near the maximum of $\R(\La)$, we can extract
 the value of $\R_M$ from the
properties of the appropriate averages. We have two different possibilities to perform 
this last step:
\begin{itemize}
 \item 
We can use the formula 
 \be \R_M^2 = - \lim_{\beta \to \infty} \lim_{n \to - \infty} {\ln
(\ba{Z_\La^n}) \over n \beta } .
\ee

We need to compute the average partition function of $n$ replicas of the models in the slightly unusual limit $n \to -\infty$.  Although there are techniques to perform this
computation \cite{PR} it would be more natural to investigate if the r.h.s. of this last equation can be computed in the limit where the dimension $D$ goes to infinity using the
standard replica method that works very well in the case $n=0$.

\item

 We can alternatively compute the moments of $K(R)_\La$, this quantity being defined in eq. (\ref{SR}):
\be
K^{(s)}=\ba{K(R)_\La ^s}\equiv \sum_{k=0,\infty} P(k,R)k^s\ .
\ee
The function $P(k,R)$ is the probability for a lattice (randomly chosen with the measure
$d\mu(\La)$) of having $k$ points inside a sphere of radius
 $R$. As far as the origin ($n=0$) is excluded, the function $P(k,R)$ is different from zero only for
even $k$ (if a point of the lattice is inside a sphere also it opposite is inside the same 
sphere). 
When $R>\R_M$ there are no lattices with no points inside a sphere of 
radius $R$. It is evident that
  \bea
P(0,R)>0 \for R<\R_M,\\
P(0,R)=0 \for R>\R_M.
\eea
The program consists in reconstructing the function $P(k,R)$ from its moments: in this way one can find
the value of $\R_M$.
\end{itemize}

 Before going on with these computations, we must chose the measure $d\mu(\La)$.
Three possibilities come to our mind. All of them are reasonable, however, depending on 
the technique we use, for some choices
the computation of the average over $\La$ may be easier.
\begin{itemize}
\item There is a natural definition (due to Siegel \cite{SIE}) of the measure over 
unimodular matrices, restricted to the fundamental region, the fundamental region being 
defined in such a way that each lattice may be represented in one and only one way by a 
matrix in the fundamental region.  Computations with this measure are technically rather 
difficult and they will not considered here.
\item
We can consider the set of matrices $\La$ (introduced by Roger) that at fixed  $\om$ depend on 
$D-1$ $\al$-variables. The action of such a matrix on a vector is defined as
\bea
(\La n)_i = \om n_i \for i<D, \\
(\La n)_D = \eta \left(n_D+\sum_{i=1,D}\al_{i}n_{i}\right) , \label{ROGER}
\eea
where the unimodularity condition implies that 
\be
\om^{D-1} \eta=1.
\ee
The $\om$-dependent measure is obtained by taking a flat measure in the interval 0-1 for
each for each of  the $D-1$ $\al$-variables. Eventually the limit $\om \to 0$ is 
taken. It is already non-trivial to prove that each lattice may be represented under the 
previous form. The proofs of this and other difficult points can be found on the paper of Roger
\cite{ROGER} and they will not reproduced here.

The computations with the Roger measure are much simpler (many results are known). Luckily enough the
two measure are equivalent for our purposes: the quantities
\be
\ba{\lan f \ran_{\La}^m },
\ee
that will play a crucial role in our study, are the same if evaluated with the Roger measure or with the
Siegel measure.

\item
As suggested by Kurchan and M\'ezard we can consider the Gaussian measure.
\be
d\mu(\La) \propto \prod_{i,k} d\La_{i,k} \exp(- \g \Tr (\La \La^*)) \delta(\det(\La)-1). \label{GAUSS}
 \ee
 where $\g$ is an arbitrary parameter. 
In this case the some of the computations can be done with the same techniques used in 
many physical problems, e.g. spin glasses \cite{BOOK}. The same measure may be useful in numerical simulations.
 \end{itemize}

As we shall see later, a simple computation shows that in the case of the Roger measure we 
have that 
\be
\ba{\lan f \ran } = \int d^Dx f(x)\ .\label{MAGIG1}
\ee

If we apply this result to the case $f(x)= \theta(R-|x|)$ we find that 
\be
\ba{K(R)}\equiv \sum_{k=0,\infty}P(k,R) k = V_{D}(R)\ ,
\ee
where $ V_{D}(R) \equiv {R_c^D \pi^{D/2}}\Gamma(D/2)^{-1}$ is the volume of the $D$-dimensional sphere  of radius $R$, 
that obviously depends also on $D$. If we
remember that the function $S(R)_\La$ may take only even integer values, we find that $P(0,R)$
must be different from zero in the region where 
\be
V_{D}(R) <2,
\ee
that is a consequence of the celebrated Minkowski theorem in large dimensions \footnote{A more  precise 
consequence of the Minkowski theorem is that $P(0,R)$ is different from zero when
$\zeta(D)V_{D}(R) <2$
where $\zeta(D)$ is the Riemann zeta function ($\zeta(\infty)=1$). As a consequence  we have that
$\R_{M}\ge R_{D}^{*}$ where $\zeta(D)V_{D}(R_{D}^{*}) =2$.}.

The proof is rather simple. Indeed using eq. (\ref{MAGIG1}) we have that:
\be
 V_{D}(R)= \ba{K(R)}= \sum_{k=2,\infty}P(k,R) k \ge 
 2 \sum_{k=2,\infty}P(k,R)=2(1-P(0)).
\ee

Let us call $R_c$ the $D$-dependent value of $R$  such that 
\be
V_{D}(R_c)=1,
\ee
its value being given by the condition
\be
\frac{R_c^D \pi^{D/2}}{\Gamma(D/2)}=1 \ .
\ee
It is convenient to measure all the quantities in units of $R_{c}$ 
when the dimension $D$ go to infinity. At this end we define:
\be
r= R/R_c, \ \ \ r_M=\R_M/R_c \ . \label{DEFI}
\ee

The problem we face consist in finding the limit to infinite dimensions $D$ of $r_M$ 
that is supposed to exist. The following
bounds are known for large $D$:
\be
1\le r_M \le 1.322.
\ee
The lower bound is the Minkowski theorem while the upper bound (the Kabatiansky- Levenshtein bound), comes from a totally different
approach that we cannot discuss in details for reasons of space \cite{SPACE}.
A simpler upper bound has derived by Levenshtein \cite{LEV}:
\be
r_M \le \frac{e}{2}=1.3591 \ .
\ee
This bound is weaker of the previous, but the proof is more direct. It is still simpler to prove the Roger bound \cite{ROGER}:
\be
r_M \le \sqrt{2}=1.414 \ .
\ee

 \section{Some known results}
 
\subsection{Roger's main theorem}
The aim of this section is to recall some known results of the value of $\ba{f^s}$ using 
the Roger measure in the case where $n<D$.  We shall see later how this annoying 
constraint ($n<D$) may be removed.

Following eq. (\ref{ROGER}) let us write $\La n\equiv X$ as using equation (\ref{ROGER}). Using equation  \form{MAGIG1} we find that
\be
\sum_{n}f(\Lambda n)=\int_{-\infty}^{\infty} d^{d}x f(x)\ .
\ee
%
%

We segue into the computations of the  higher moments
of $f$.
%
Fortunately the appropriate
computations have been done in rigorously way by Roger. One finds that his results are relatively
simpler in position space.

The strategy for controlling the previous formula consists of a few steps:
\begin{itemize}
	\item We classify all sets of $s$ vectors $n$ according to their linear 
dependence.
\item We perform the average and the sums inside each class.
\item At the end we write 
the sum over all possible classes.
\end{itemize}
The crucial theorem is based on the following two lemmas.
\begin{itemize}
\item
One can prove
\be
\ba{\prod_{k=1,s} (\sum_{n^k_i} f(\La x^k))}=\left(\int d^Dx f(x) \right)^s, 
\ee
where the sum is done over all sets of $k$ vectors $ n^k$ belonging to $Z^D$ that are linearly
independent (as usual the origin, i.e. $n^{k}=0$, never appears). The formula is valid only for $s\le D$.
(It is obvious that we cannot find more than $D$ linearly independent vectors in 
dimensions $D$.)
This formula has a very simple meaning. Each of the $s$ points $ n^k$ may be carried in any point
of the space independently on the other one provided that they are linearly independent.

\item

In order to obtain the final result need also to consider the average of similar sums 
restricted to the case where the vectors $ n$ are linearly dependent.

More precisely we introduce $s$ vectors $ n$ that are linear combination of $s-h$ linear 
independent vectors and their linear dependence is specified by a matrix $M$.  In other 
words the $s$ vectors $ n$ span a $(s-h)D$ dimensional space and they satisfy the $h$ 
different linear conditions:
\be
\sum_{k=1,s} M_{j,k} n^k_i =0 \for j=1,h \label{CON} \ ,
\ee
where the matrix $M$ has integer elements and it is irreducible (there is no integer 
matrix $M'$ such that $M'=pM$ with $p$ integer).  In this case we find that
\be
\ba{\prod_{k=1,s} \left(\sum_{n^k_i}^{(M)} f(\La n^k)\right)}= \N(M) \int\prod_{k=1,s} d^Dx^k f(x^k)
\prod_{j=1,h} \de^D\left(\sum_{k=1,m} M_{j,k}  x^k\right)\, ,
\ee
where the sum is done on vectors that satisfy the condition (\ref{CON}) (that depend on 
the matrix $M$ and have $s-h$ linear
independent components). The quantity $\N$ is a normalization factor that 
is equal to 1 in many cases and that for our purposes may be taken  equal to 1.
 \end{itemize}
We now we all the elements to implement the strategy to compute the moments. We must  
consider all possible linear dependence of the vectors $n$ and we must reduce the sum over all 
the values of $n$ to a sum over all possible linear dependencies.

We can now quote the main theorem of Roger \cite{ROGER}:
\be
\ba{<f>^s}=\sum_{h=0,s-1} \sum_M \N(M) \int\prod_{k=1,s} d^Dx^k f(x^k)
\prod_{j=1,h} \de^D\left(\sum_{k=1,s} M_{j,k} \vec x^k\right) \, , \label{ROGEREQ}
\ee
where the sum is done over all the sets of $ h \times s$ matrices $M$  
corresponding to different linear conditions \footnote{In order to verify if two matrices $M$ correspond to  different linear conditions, Roger  writes the
matrices $M$ in a {\it canonical} form and presents his main theorem in this form. However we do not
need his theorem under this form.}.

\subsection{The case $s=2$}
Let us show how this strategy works in the case $s=2$.
Here we have to compute
\be
\ba{\sum_{n_1,n_2}f(n_1)f(n_2)}\ .
\ee
Now we have two possibilities:
\begin{itemize}
\item 
The two vectors $n$ are linearly independent. We obtain the following contribution
\be
\int d^Dx_1 d^Dx_2 f(x_1) f(x_2) \ .
\ee
\item
We have to consider the case where the two vectors $n$ are linearly dependent.  In this 
case we can write the constraint in an unique way as \be q_1  n^1 + q_2  n^2 =0,
\ee
with if we restrict ourselves to the case of positive $q_{1}$ and 
$(q_{1},q_2)=1$ (i.e. the pair $q_1$ and $q_2$ is irreducible)
\footnote{The expression $(a,b)$ denotes as usually the maximum common divisor of $a$
and $b$.}.
We finally finds the following contribution
\be
 \int d^Dx d^Dy f(x) f(y) \de^D(q_1  x +q_2  y) = \int d^Dx 
 f(q_1 x) f(q_2 x ).
\ee 

Putting everything together we find the final expression
\be
\ba{\sum_{n_1,n_2}f(n_1)f(n_2)}
= \int d^Dx_1 d^Dx_2 f(x_1) f(x_2)+\sum^{I}_{q_1,q_2} \int d^Dx f(q_1 x) f(q_2 x ), \label{S2}
\ee
where the sum $\sum^I$ is restricted over the irreducible pairs with  
$q_1$ positive.
\end{itemize}

We have already seen how this construction works for $s=2$, in the next subsection we will consider
in details the case $s=3$.

 \subsection{ The case $s=3$}
If we apply the previous formula in this case we have only three possibilities: $h=0$, $h=1$ and $h=2$ 
($h$ being the number of linear constraints).
\begin{itemize}
\item $h=0$ 
In this case the matrix $M$ does not exist and the corresponding  contribution is simply given by
\be
\int d^Dx_1 d^Dx_2 d^Dx_3 f(x_1) f(x_2) f(x_3) \ .
\ee
\item $h=1$ 
In this case we have a linear constraint of the form
\be
q_1x_1+q_2x_2+q_3x_3=0\ .
\ee
We have two possibilities.
\begin{itemize}
\item 
One of the three $q$ is equal to zero.
 Here we have three equal contribution. If we set $q_3=0$, we
must impose  that $q_1$ and $q_2$ have no common factors (i.e. $(q_1,q_2)=1$).
We finally find a contribution equal to
\be
3 \sum_{q_1,q_2}  \int d^Dx_1 f(q_1 x_1) f({q_2 x_1} ) \int d^Dx_3 f(x_3) \ .
\ee 

\item All the $q$ are non zero.
The previous condition requires that $q_1$, $q_2$ and $q_3$ have no common factors.
We finally find
\be
\sum^I_{q_1,q_2,q_3}  q_3^{-D} \int d^Dx_1 f(x_1)  \int d^Dx_2 f(x_2)f\left({q_1 x_1 +q_2
x_2\over q_3} \right) .
\ee 

\end{itemize}
\item $h=2$
In this case  we can write the two linear constraints as  
\be
q_1 x_1 =q_2 x_2\ \  \mbox{and}\ \ g_1 x_1 =g_2 x_3.
\ee
We finally finds that the contribution from this case is given by
\be
\sum^I_{q_1,q_2,g_1,g_2} (q_2 g_2)^{-D} \int d^Dx f(x) f\left({q_1 x\over q_2} \right)f\left({g_1 x\over 
g_2} \right) \ .
\ee 

\end{itemize}
The final result is given by the sum of all the previous contributions.

In the same way we can write explicit formulae that contain more and more terms when $s$ 
increases.

\section{Some simple computations}

In this section we will present some simple computations, in order to familiarizes 
ourselves with the previous results.

If we look to the previous formulae it is evident that there is an obvious  
prefactor that  goes to zero very fast when $D\to \infty$ for many choices of the 
function $f$.  Let us see what happens in a some simple examples and/or doing some simple 
approximations.

\subsection{The Gaussian case}

Let us consider  the case where the function $f(x)$ is given by 
\be
f(x) = (2 \pi )^{-D/2} \exp(-\beta x^2/2).
\ee
Here the integrals can be easily done and we finds that
\bea
<f> = \beta^{-D/2} \\
<f^2> = \beta^{-D}+ \beta^{-D/2} \sum^I_{q_1,q_2} (q_1^2+q_2^2)^{-D/2} \\
<f^3>= \beta^{-3D/2} + 3 \beta^{-D} \sum^I_{q_1,q_2} (q_1^2+q_2^2)^{-D/2} + \nonumber \\
\beta^{-D} \sum^I_{q_1,q_2,q_3}  (q_1^2+q_2^2+q_3^2)^{D/2} 
+\beta^{-D/2} \sum^I_{q_1,q_2,g_1,g_2}  (q_1^2 g_1^2+q_1^2 g_2^2 + q_2^2 g_1^2)^{-D/2} \nonumber 
\eea

For large $D$ we find:
\bea
\frac{2 \ln (<f>)}{D} = \beta^{-1} \\
\frac{2 \ln (<f^{2}>)}{D}\approx \max(\beta^{-2}, (2\beta)^{-1})\\
\frac{2 \ln (<f^{3}>)}{D}\approx \max(\beta^{-3}, (2\beta^{2})^{-1},(3\beta)^{-1}) \nonumber
\eea

The  problems connected with the limit $D\to \infty$ are clear from the previous formulae.
Depending from the value of $\beta$ different terms are the leading ones (the case 
$\beta>1$ and $\beta<2$ are quite different). Sometimes it happens that
the leading term from one contribution is smaller of the subleading terms of other contribution.
The terms with the $q$'s and the $g$'s equal to 1  are always the leading ones inside a given class.

\subsection{The theta function}

We consider here the case of the function:
\be
f(x) =\theta (R-x). \label{THETA}
\ee

Our plan is to compute the leading term of each moment of $f$ when the dimension goes to infinity, 
resuming the resulting series and extracting information on the  probability $P(k,d)$. The
relation $f^2(x)=f(x)$ will be useful in simplifying the result. 

Let us use the previous formulae for computing the first three moments.
We find in the same way as before that
\bea
K^{(1)}=\ba{<f>} = V_{D}(R), \\
K^{(2)}=\ba{<f>^2} = V_{D}(R)^2+ V_{D}(R) \sum^I_{q_1,q_2} (\max(q_1,|q_2|)^{-D}  \\
K^{(3)}=\ba{<f>^3}= \nonumber \\ V_{D}(R)^3 +  V_{D}(R) \sum^I_{q_1,q_2} \max(q_1,|q_2|)^{-D}+ 
 V_{D}(R)^2 \sum^I_{q_1,q_2,g_1,g_2} ( {\max(1,|q_2/q_1|,|g_2/g_1| \over q_1 q_2})^{-D}+\nonumber \\
\sum^I_{q_1,q_2,q_3} (q_3)^{-D} \int d^Dx_1   \int d^Dx_2 (\theta(z_1-R)\theta(x_2-R)
\theta(|q_1x_1+q_2x_2|-q_3 R) \nonumber
\eea

We have terms that scale as different powers of $V_{D}(R)$.  Inside each class the leading 
terms are those the variable $q$ and $g$ are equal to $\pm 1$.  If we keep only these 
terms we find the simple result
\bea
K^{(1)}=<f> = V_{D}(R) \\
K^{(2)}=<f^2> = V_{D}(R)^2+  2 V_{D}(R)   \\
K^{(3)}=<f^3>= \nonumber \\ V_{D}(R)^3 + 2 V_{D}(R)^2 
 +4 V_{D}(R) +4 \int d^Dx_1   \int d^Dx_2 \theta(x_1-R)\theta(x_2-R)
\theta(|x_1+x_2|-R) \nonumber
\eea

The last integral is exponentially small for large $D$ with respect to the other terms (see  also the
appendix): it is  the probability of having three overlapping spheres
of the same radius. This crucial results follow from two general facts \footnote{The argument is well known and plays a crucial role in the evaluation of the leading term of the
Mayer expansion for an hard sphere gas in infinite dimensions (see \cite{PS}).} that are valid when $D\to \infty$:
\begin{itemize}
\item
The measure on a $D$ dimensional sphere is concentrated on its surface.
\item
Two generic vectors (of length 1) have vanishing scalar product; therefore the sum of two generic vectors of
length less or equal to $R$ is at most $2^{1/2} R$.
\end{itemize}

Also in this case (depending if $R$ is larger or smaller that $R_{c}$), different terms 
dominates the result for each moment when the dimension becomes large.
           
\subsection{Summing the leading terms}\label{SUM}

Our aim it to obtain a closed formula for the leading terms for each moment of the 
function $f$ in order to extract the asymptotic behaviour.  Unfortunately controlling the 
leading term in each moment does not imply that we control the leading term in the 
probability distribution.

 From the analysis of the first three moments   we conclude that in
general the terms that are leading when the dimension goes to infinity are those that come from
 $h$ linear conditions of the form
\be
x_i=\pm x_j.
\ee

If we select these contributions we find that the result can be given in terms of $h$
integrals of the same function. In the case of a generic even function 
(i.e. $f(-x)=f(x)$) we get:
 \be
K^{(s)}=\ba{<f>^s} \approx \sum_{h=1,s}  \sum_{\nu_i=1,s} C(s,h,\nu) \prod_{i=1,s} 2^{\nu_i-1} \int dx^D
f(x)^{\nu_i} \ee
where the sum is done over all the sets of $h$ integers $\nu_i$ such that 
\be
\sum_i {\nu_i} = s,
\ee
and $C(s,h,\nu)$ is a crucial combinatorial factor that is equal $h!^{-1}$ times the number of ways
in which we can divide $s$ objects in $h$ group of  $\nu_i$ elements (we have to specify which 
variables  are equal). 
This number is given by
\be
C(s,h,\nu)={ s! \over h! \prod_{i=1,h} \nu_i!}\ .
\ee

For example in the case $s=3$ and $h=2$, we two equal contributions 
\bea
\nu_1 = 2  \ \ \  \nu_2=1 \\
\nu_2 = 2  \ \ \  \nu_1=1 
\eea
that correspond to the same term (the number of contributions compensate the term $m!^{-1} 
=1/2$). The remaining factor 3
correspond to the three contributions:
\be
\int dx_1 dx_2 dx_3 [\de(x_2-x_1)+\de(x_3-x_1)+\de(x_3-x_2)] \ .
\ee

The formulae simplify if we compute the generating function of the moments. We find after some simple
algebra that
\be
\ba{\exp y <f>} \approx \exp ({1 \over 2} \int dx (\exp ( 2 y f(x)) -1)) \ .
\ee

If we apply this formula to the case of the function in \form{THETA} we find that
\bea
\ba{\exp y K(R)_\La} \approx \exp \left(-{V_{D}(R) \over 2}  \exp( 2 y  -1)\right)=\\
\exp \left(-{V_{D}(R) \over 2}\right)\sum_{k=0,\infty} {1 \over k!} \left({V_{D}(R) \over 2}\right)^k \exp (2 k y)
\eea

We finally find the simple result:
\be
P(2k,R) \approx \exp \left({-V(R) \over 2}\right) {1 \over k!} \left({V(R) \over 2}\right)^k,
\ee 
while $P(2k+1,R)=0$ (The number of points of a lattice inside a sphere is alway even!).

 As a byproduct
of this Poisson distribution we have that
\be
P(0,R) \approx \exp \left(-{V(R) \over 2}\right)\ .
\ee

Let us  find the consequences of this approximate result. We have called $R_c$ the radius of a sphere
of unit volume. If we take a large value of $D$ at fixed ratio 
$r\equiv {(R / R_c)}$ 
we find that $V(r R_c)$ goes to zero or to infinity depending if $r$ is smaller or greater than $1$.
\begin{itemize}
\item
If $r<1$ we have that $P(0,rR_c)$ goes to 1 apart from exponentially small corrections. With
probability going to 1  lattices do not have points at distance smaller that $R_c$. 
\item
In the interesting case $r>1$   this computation give an exponentially small but non vanishing
contribution for $P(0,rR_c)$. 
\end{itemize}
 If we stick to this result we obtain that $P(0,rR_c)$ is always non zero and that we can find always
a lattice that contains no points at distance less than $r R_c$. This result cannot be correct
because it can be proved that no such lattice exist for large $D$ as soon 
$r>1.32$.  

This failure is
due to the fact that we have neglected exponentially small terms. This  is allowed only if these
terms do not compensate the exponential small term we have obtained for the probability 
distribution  and produce  a net zero result. In other words we cannot anymore neglect the 
subleading terms if the sum of the of the leading terms becomes smaller of each individual 
subleading term.
It was proved by Roger that this does not happens in the relatively smaller region where
$r^D<D$ (a similar region appears in the analysis of \cite{PZ}).

Up to now we have transcribed the results of Roger in a slightly different language skipping the
proof  of his theorems.  Our aim is to extend our command of the expression for $P(k,R)$ by
considering more terms in such a way to be able to control better the remaining integrals. This
will be done in the next sections.

\subsection{The appearance of the random energy model}\label{REM}
Before going to more precise computations, we it is convenient to  study of also the 
partition function if we retain only the the terms considered in the previous subsection.
We recall the definition of the partition function given in eq. (\ref{PART}).
\be
Z(\beta)_\La \equiv \lan\exp ( -\beta |x|^2)  \ran_\La \ .
\ee

If we apply the formulae of the previous subsections we find after some algebra that
\be
\ba{Z(\beta)^s}=  2^s\sum_{h=1,s}  \sum_{n_i=1,s} C(s,h,n) \prod_{i=1,h} 2^{-1}
 \left({2 \pi \over \beta n_i}\right)^{-D/2} \ , \label{PARTEQ}
\ee
where (as usual) the sum is done over all the sets of $h$ integers $n_i$ such that $\sum_i {n_i} =
s$.

The previous formula does not look particularly illuminating, however a close look 
shows that this   formula is well known to people studying the random energy model (REM) 
\cite{DERRIDA} .
In this way we can rewrite it in a more illuminating way.  At this end let us 
consider a new model where the partition function can be written as
\be
Z_R(\beta)= \sum_{k=1,N} \exp( -\beta E_k) \ ,
\ee
where the $E_k$ are  $N$ random independent quantities distributed with the probability
distribution $P^{(N)}(E)$.
Standard arguments shows that
\be
\ba{Z_R^s}= \sum_{m=1,s}  \sum_{n_i=1,s} C(s,m,n) \prod_{i=1,m} p^{(N)}_{n_i} \, ,
\ee
where 
\be
p^{(N)}_{n}= {\int dE P^{(N)}(E) \exp (-n \beta E) \over N} = \int dE p^{(N)}(E) \exp (-n \beta  E)
\ee
and we have defined 
\be
p^{(N)}(E)={P^{(N)}(E) \over N}\ .
\ee
In other words $p^{(N)}(E)dE$ is the probability of finding an energy in the interval 
$[E,E+dE]$.

The limit $N$ going to infinity can be done without difficulties if the function $p^{(N)}(E)$
depends on $N$ in such a way that this function remains finite in this limit,i .e. if the following limit exists. 
\be
p(E) = \lim_{N\to \infty} p^{(N)}(E)
\ee
The normalization condition
($\int p^{(N)}(E) dE =N$) implies that
\be
\int p(E) = \infty,
\ee

In our case we have to take
\begin{equation}
p(E) = \lim_{N\to \infty} p^{(N)}(E) = \frac{D}{4} E^{(D-1) / 2}V_{D}
\end{equation}

It is immediate that within the approximation of summing the leading terms the partition 
of our model coincide (if we neglect the factor $2^s$ in eq.  (\ref{PARTEQ}) \footnote{The 
factor $2^s$, that is irrelevant for most conclusions may be easily obtained by adding a 
fixed degeneracy (i.e. 2) of the levels.} ) with a random energy model where
\be
p_n = 2^{n-1}\left({2 \pi \over \beta n_i}\right)^{-D/2} =\frac{1}{2} \int d^Dx \exp (-n \beta x^2/2)
\ee
and therefore
\begin{equation}
p(E)= \frac{1}{2} \int d^D x \de(E- x^2/2) \ .
\end{equation}

In other words we consider $N$ points $x_{k}$ in the $D$-dimensional space that are 
uniformly randomly distributed with density $1/2$ inside a sphere of an appropriate radius going to infinity with $N$.
The partition function is just the limit $N \to \infty$ of 
\be
Z_{R}(\beta) =\sum_{k=1,N} \exp (-\beta x_{k}^{2}/2) \ .
\ee
This implies that the probability distribution of the quantities $|\lambda n|^{2}$ 
correspond to a flat uncorrelated distribution of the points $\lambda n$. The probability 
of finding an arbitrary large sphere of the space with no points, is always non zero although it 
is very small for a quite large sphere. this is perfectly consistent with the results of 
the previous subsection, where it was shown that there is no upper limit for the radius.

\subsection{A detailed analysis of the Random Energy Model}
With this identification we can proceed to a direct analysis of the REM.  Within this approximation we would like 
to follow the strategy outlined in section II, where we connect the maximal radius 
with the behaviour of $ \ba{Z(\beta)^{s}}$ in the double limit $ \beta \to \infty$ and 
$s \to -\infty$.

The first method we could use consists in computing the moments of the partition function in the REM replica
method. In the case of the REM the replica method is well established in this case and it gives the same results of more 
conventional methods.

Let us firstly follow this route. In evaluating the l.h.s. of eq. (\ref{PARTEQ}) we
consider the contribution where all the $n_i$ are equal to $u$. In
this case their number is $k=s/u$. This contribution gives
\begin{eqnarray} \nonumber
\ba{Z(\beta)^s}\approx 2^s s!/(u!)^k \exp \left(-s{D/2 \ln(\beta u /(2 \pi) +\ln (2) \over u}\right)\\
= 2^s s!/(u!)^k \left( {\int d^D x \exp(- u \beta x^2/2) \over 2}\right)^{s/u}
\end{eqnarray}

The previous formula can be used as input to the replica method. After some computations
(that we skip) we can try to use the appropriate saddle point method in the limit $D \to \infty$. 
Within this approximation, neglecting those prefactors that have a constant limit
when $D \to \infty$, we find that in the low temperature region (i.e. high beta) the 
moment of the partition function are given by:
\be
\ba{Z(\beta)^{s}}\approx \exp( - A D s \beta ), \label{PREVIOUS}
\ee
where the constant $A$ is given by
\be
A=\pi e
\ee

In other words we find that all the realization give the same partition function with probability one.
Unfortunately we are interested to the behavior the partition function in the tail of the 
probability distribution and this require the study of the limit
of large $\beta$ and $-s$. In this situation the prefactors cannot be neglected and the analysis
becomes more complicated.

If we look to the formulae of the REM we find that for sufficiently high $\beta$ we have that the
partition function is dominated by the ground state and that the probability distribution of the
minimal energy is given by
by
\be
\PP(E) =p(E) \exp (-\int_{e<E} de\  P(e)) \propto E^{D/2} \exp( - V_{D}/2 E^{D/2}).
\ee
We thus find that
\bea
\ba{(Z/2)^s} \approx  \int dE \PP(E) \exp(-\beta E)\propto \\
 \int d^Dx \exp(-\beta s x^2/2 -\frac12 V_{D}x^D) \propto \int dR^{D} \exp ( -\frac{1}{2} V_{D}R^D- 
 \frac{1}{2} s \beta R^2)
\eea
If $D$ goes to infinity first we find the previous result. On the other hand, if we have a large value
of $\beta$ for negative $s$ at fixed dimension, neglecting the prefactors we find that 
\be
\ba{Z(\beta)^s} \approx \exp \left( -B(D) (\beta s)^{D \over D-2}\right) \ . 
\ee
where 
\be
B(D)={d-2 \over D}\left(DV_{D}\right)^{-2\over D-2}\approx_{D \to \infty} \pi e
\ee

The problems arising in exchanging the two limits $D$ and $\beta$ to infinity for negative $s$ are
clear from the previous formulae. Moreover within our approximation the limit
\be
\lim_{s\to - \infty} \frac{\ln(\ba{Z(\beta)^s})}{s}
\ee
does not exist, contrary to our expectations. This negative result reflect our lack of 
command of the tail of the partition function within our approximation.

These computations clearly shows that the approximation of keeping only the leading term 
is not enough to find the most compact regular lattice and that a more refined 
approximation is needed. This will be the subject of the next sections, where we are 
going to present new results and conjectures (up to now we have only translated in a 
different language  the results of Roger \cite{ROGER}).

\section{The importance of being connected}
\subsection{The introduction of connected moments}
Before considering further contributions we need simplify the previous computation.
At this end it is important to introduce the connected moments
and to do  some general consideration.

 Let us consider an integer valued function $f$ and 
its moments defined as
\be
<f^s>= \sum_k P(k) k^s\equiv f_s \ ,
\ee
where $P(k)$ is the probability that the function $f$ is equal to $k$.

It is usual to introduce the generating function 
\be
G(z)=\sum_{s=0,\infty}f_s \frac{z^s}{s!} = \sum_{s=0,\infty}\sum_{k=0,\infty} P(k) 
\frac{(zk)^s}{s!}=
\sum_{k=0,\infty} P(k) \exp(zk)=  \exp(C(z)) \, , 
\ee
where $C(z)$ is the generating function of the connected moments:
\be
C(z)=\sum_{s=1,\infty}<f^s>_c z^s \ .
\ee
Let us assume that we can write 
\be
C(z)= B \sum_{k=1,\infty} p(k) (\exp(zk) -1),
\ee
where the quantities $p(k)$ are non-negative and such that
\be
\sum_{k=0,\infty} p(k)=1.
\ee

Using the theorem of composed probability we readily find that 
\be
P(k)=
\exp(-B)\sum_{n=0,\infty}{B^{n} \over n!}\prod_{i=1,n}\left(\sum_{k_i=1,\infty} p(k_i)\right) 
\de(\sum_{i=1,n} k_i-k) 
 \ee
In other words the distribution $P(k)$ can be obtained in the following procedure:
\begin{itemize}
\item
We first extract a number $n$ with a Poisson distribution with average $B$.
\item 
We extract $n$ independent numbers $k_i$ with probability $=p(k)$.
\item
The quantity $k$ is given by $\sum_{i=1,n}k_i$.
\end{itemize}

We finally find that 
\be
P(0)= \exp(-A) =\exp(-\lim_{z\to-\infty}C(z)).
\ee

We see that the problem of computing the value of $P(0)$ is reduced to the evaluation of the
generating function of the connected moments in a particular limit. 

On the other hand, let us assume that there exist
an analytic function $ f_c(s)$ such that for positive integer $s$
\be
 f_c(s)=<f^s>_c,
\ee
and the function $ f_{c}(s)$ does not have a nasty behavior at infinity in the complex plane. If this function  does
 not have singularities in the region $\Re e \ s \ge 0$ one can prove  that
 \be
\lim_{z\to-\infty}C(z) = f_c(0).
\ee
Under the previous assumptions the task of computing $P(0)$ is reduced to the problem of 
finding an 
analytic expression for the function $f_c(s)$ and of computing it at $s=0$.  Our strategy 
will consists in summing subsets of all the possible contributions and to evaluate the 
result at $s=0$ \footnote{The reader may be  puzzled  by this result that  has a 
strong replica flavour.}.

If we use the  computation of the moments done in the previous section, where only the 
leading terms where taken into account, we find the very simple result for the connected 
moments: 
\be
 f_c(s)= \int d^Dx f(x)^s \ .
\ee

 If we take the function $f$ equal to $K(x)\equiv \theta(R-x)$ we find that
\be
\tilde f(s)= \int d^Dx K(x)^s =V_D
\ee
that is obviously an analytic function of $s$ in the whole complex plane.

\subsection{How to compute the connected moments}
It is evident from the previous discussion that the direct computation of the connected 
moments is extremely interesting.  The final expression for the connected moments 
certainly contains less terms than the expression for the moments: this is not a surprise 
to people working in statistical mechanics.

We discuss now how this task may be achieved. 
Let us consider what happens up to the third moment. If we start from the formulae of the  section III the same notation,
we find:
\bea
<f>_c=\int d^Dx f(x)\\
<f^2>_c=\sum^R_{q_1,q_2} \int d^Dx_{1} d^Dx_{2} f(x_{1}) f(x_{2} ) 
\delta(x_{1}q_{1}+x_{2}q_{2})\\
<f^3>_c=\sum^R_{q_1,q_2,q_3}  \int d^Dx_1    d^Dx_2 d^Dx_3 f(x_1) 
f(x_2)f(x_3) \delta (q_1 x_1
+q_2 x_2 + q_3 x_{3} ) +\\
\sum_{q_{1},q_{2},g_1,g_2}^{R} \int d^Dx_1    d^Dx_2 d^Dx_3  f(x_1) 
f(x_2)f(x_3)  \delta (g_1 x_{1}+g_2 x_{2}) \delta (q_1 x_{1}+q_2 x_{3})
\eea
We ready see that the only terms that remains in this limit are connected (in the same 
way of connected diagrams) in the sense that they
cannot be split into the products of terms that contain integrals that can be done 
independently.

In order to see how derive a similar result for the higher moments 
and to stream down the discussion let us  set
all factor $\N / q$ to 1. In this case the general formula of section III can be written as
\be
<f^s>= \sum_{G}\int \prod_{i=1,n} (d^Dx_i f(x_i))  G(\{x\}) \ ,
\ee
where the sum is done over the appropriate set of function $G(\{x\})$. These functions are
 an appropriate product of delta functions that enforces the linear dependence of
some of the $x$'s.

We state that a function $G$ is connected if it goes to zero when one or more of the $x$ go to
infinity together. In other words we must have that
\be
\lim_{\la \to \infty} G^n_k(x_1+\la c_1,x_2+\la c_2,....,x_n+\la c_n) =0
\ee
for any set of vectors $c$ such that at least one $c$ is different from zero and no pairs of  $c$ 
are  equal.

If $G$ is not connected it can be written as the product of its connected pieces.  if now 
we take care of all the multiplicity factors and we use the standard manipulations of 
statistical mechanics we find that
\be
<f^s>_{c}= \sum_{G_{c}}\int \prod_{i=1,n} (d^Dx_i f(x_i))  G_{c}(\{x\})
\ee
where the sum is done only over those $G$ that are connected (we call them $G_{c}(\{x\})$) \footnote
{Some people may consider the previous formula nearly self evident. Indeed let us consider a 
function $f$ that is the characteristic function of a domain of volume $V$. 
General considerations may be used to argue that $<f^s>_{c}$ should be proportional to $V$ 
and this is possible only if only connected $G$ are present in the r.h.s of the previous 
equation.}.
%

\section{A crucial conjecture}
\subsection{A serious difficulty}

Before going further we must face a problem that we have postponed up to now.

The program we have put forward  may seem witless. We have 
said that we can write equations of the form
\be
\ba{f^s}= F(s,D) \, ,
\ee
that are valid only for $s < D$ (as we have seen the r.h.s of the previous equations 
contains some sums over integers and some integrals).  Roger theorems (that are at the 
basis of our computations) does not give useful information for $m>D$ for the very good 
reason that in this case both sides of equation \form{ROGEREQ} are infinite.  Indeed there 
are very asymmetric lattices for which the function $<f>_\La$ is very large (e.g. there are 
many points very near to the origin). Conversely the  
the sum over the $M$ is not always valid. Indeed it is easy to check that the 
formula \form{S2} for $\ba{f^2}$ is divergent in two dimensions as soon as $f(0)\ne 0$.

How can we write a meaningfully expression in finite dimension for $\ba{\exp(y <f>}$ if we 
can compute only a few moments of the $<f>$?  We would like to put forward a reasonable 
conjecture that should allow us to obtain an explicit form it.  Before doing it, let us 
look better to the previous formulae and see what happens if we refine the previous 
computation by including some additional terms.

\subsection{A more refined computation}

Let us consider for simplicity only the connected moments. The non-connected one can be
obtained from them by combinatorial factors. We want now to analyze in details the origin of the
divergences that we have seen in the previous subsection.

Let us start by considering the second moment in the case where the function is equal to 
$\exp (- x^2)$.  In this case we have  explicit formulae. Doing the appropriate integrals 
on obtain that  
\be
<f^{2}>_{c}=\sum^{I}_{q_{1},q_{2}}\left( {\pi \over q_{1}^{2}+q_{2}^{2}} \right) ^{D/2}
\ee
As we will see  in dimensions larger than two 
the sums are convergent.  

In the same we we obtain
\bea
<f^{3}>_{c}=\sum^{I}_{q_{1},q_{2},q_{3}}\left( {\pi^{2} \over q_{1}^{2}+q_{2}^{2}+q_{3}^{2}} 
\right) ^{D/2}\\+
\sum^{I}_{q_{1},q_{2},g_{1},g_{2}}
\left( {\pi \over q_{1}^{2}g_{1}^{2}+q_{2}^{2}g_{1}^{2}+q_{1}^{2}g_{2}^{2}} \right) 
^{D/2}
\eea

In dimensions three only the first term is divergent.  The other term is divergent in 
dimensions two.

This is a general feature.  The most divergent term is always the one that correspond 
to a single linear constraint.  We can therefore start by considering only this 
contribution.  We want to arrive to a final simple expression, that can be reached in a 
few steps.

Within the approximation of taking the linear constraint we have

\be
\ba{<f>^s}_c \approx  \sum^I_{q_1 \cdots q_s}  \int \prod_{k=1,s} dx_k^D f(x_k)
\de(\sum_{i=1,s} q_i x_i) 
\ee
where by the superscript $I$ we indicate the fact that there is no common factor among
all the $q$.
In our case we find
\begin{equation}
\ba{<f>^s}_c \approx \pi^{-D/2} \sum_{q_1 \cdots q_s}^I  
({\pi^{s-1} \over \sum_{i=1,s} q_i^2 })^{D/2} 
\end{equation}

We note that if $H$ is an homogeneous function of degree $-\nu$ 
\be
\sum_{q_1 \cdots q_s} H(q) = \zeta(\nu) \sum_{q_1 \cdots q_s}^I H(q) \ .
\ee

We finally find that (within the approximation of keeping only the linear constraint)
\be
f_c(s)={\Gamma(D/2)\over \pi^{D/2} \zeta(D)} \int {dt \over t} t^{D/2} G(t)^s,
\ee
where
\be
G(t)=\pi^{D/2}\sum_{k=1,\infty}  \exp (- t k^2 ).
\ee
Few remarks are in order.
\begin {itemize}
\item
In this case the  function $f_c(s)$  can be written under the form
\be
f_c(s)= \int dz p(z) z^s,
\ee
where
\be
p(z)\propto\int {dt \over t} t^{D/2} \de(G(t)-z).
\ee
It is crucial to note that the expression for $p(z)$ is well defined also in the case 
where some of the moments of the function $f$ are divergent. The divergence of the moments 
of the function $f$ is related to the asymptotic behaviour of the function $p(z)$ at 
large $z$.
\item
We have that for small $G$ 
\be
G(t) = \left({\pi \over t}\right)^{D/2} +\tilde G(t)
\ee
where $\tilde G(t)$ is a $C^\infty$ function at $t=0^+$ (it is a well known fact that can 
be easily proved using the Poisson formula).
Therefore the function $f_c(s)$ has simple poles on the real axis at integer values of $s$ starting
from $s=3$. 
The first pole correspond to a decrease of the function $p(z)$ as $z^{-3}$ at large $z$.
\item

For dimensions $D>s$  the quantity $f_c(s)$ is an analytic function of $D$ that has some
poles for some values of $s$. Moreover $f_c(s)$ is an analytic function of both variables $s$ and $D$
\end{itemize}
\subsection{A preliminary conjecture}
We can hope that these analyticity properties in $D$ and $s$ are true in general. If we accept this
conjecture  each given moment can be computed at sufficient high dimension, where it is
convergent, and we can  evaluate it  as analytic continuation in $D$ and $s$ at the point we need it. 
The analytic continuation of the moments in the dimensions at fixed $s$ and analytic continuation of the moments in the value 
of $s$ at fixed dimension should coincide. If in a given dimension they are both 
singular, the form of the singularity in $s$ and in $D$ are related.

If this conjecture is true, Roger formulae bring information on the function $f_c(s)$ 
also in the region where the definition of the moments is divergent and therefore our program is not witless.

This conjecture is quite strong.  It is possible to verify by explicit computations that 
the $f_c(s)$ function has the needed analytic properties in the region where Roger 
formulae are convergent.  On the other hand it is not obvious that the function $f_c(s)$ 
at given dimension coincides with its analytic continuation from high dimensions.  We do 
not know how such a conjecture could be proved in a rigorous way.  It would be interesting 
to check if this conjecture is supported by a numerical study of the first moments in low 
dimensions.  We will not attempt to further study this point here.
 
 \section{A Comparison with the gas of Hard spheres}
\subsection{The hard spheres gas}
Here we try to get some tentative conclusions for the asymptotic behaviour of $r_{M}$ in 
large dimensions. In order to do this we have first to recall some properties of the gas of
hard spheres.

An object that it is quite familiar to physicist is the partition function of a gas of hard sphere of
diameter $r$.
 \be
Z(V,N)= \int \prod_{i=1,N}d x_i \prod_{i,k=1,N} \theta (|x_1-x_k|-R)
\ee
The partition function depends on the total number of particles $N$ and on the volume $V$ of the box
where the particle are confined.
Without loss of generality we can assume that $V=N$, i.e. the density is one. In the infinite volume
limit we have that
\be
Z(V,N) \approx \exp( -V \  S(r))\ ,
\ee
where $S(r)$ is the entropy density.  The entropy is a function of the reduce diameter $r$ and it 
should diverge when $r$ arrives to the point ($r_{A}$) that correspond to the highest 
density packing (for $r>r_{A}$  $Z=0$).  It is hard problem to decide if in the infinite volume 
limit the highest density packing is a lattice packing.

In dimensions $3$ the most compact packing is a lattice packing. Therefore the partition 
function vanishes as soon as $r>r_{M}$ and 
\be
\lim_{r\to (r_{M})^{-}} S(r) =-\infty
\ee
A crucial phenomenon for a gas of hard spheres is the phenomenon of crystallization: 
there is a reduced radius $r_{C}$ such that for $r>r_{C}$ the configurations of the hard 
particles are very similar to those of a regular crystal. The transition form the liquid 
to the crystal phase is a first order transition.

Generally speaking in higher dimensions we can have two possibilities
\begin{enumerate}
	\item
	The partition function of hard spheres is different from zero also in the region $r>r_{M}$ 
	and diverges at a value of $r=r_{A}>r_{M}$. 
	In this case $r_{A}$ is the maximal radius of a non-crystalline  (i.e. amorphous) 
	packing and is bigger of the maximal radius of a crystalline packing. This situation is 
	quite different from the three dimensional situation.
	
	\item 
	As in the three dimensional case the partition function of hard spheres becomes zero at 
	$r_{M}$. We can however distinguish among two cases.
	\begin{enumerate} \item
	There is a radius $r_{C}$ where the entropy has a first order transition where the system crystallize.
	\item
	There is no crystallization transition. \label{LAST}
	\end{enumerate}
	\end{enumerate}

The situation become more complex if we consider the possibility of having a glass transition in the non-crystalline phase \cite{PZ} at $r_{G}$.  This glass transition may be or in the
metastable phase above $r_{C}$, has happens in three dimensions, or in the stable liquid phase below $r_{C}$. This glass phase transition  is important because the virial expansion 
does not give information on the behaviour above $r_{C}$. The arguments of \cite{PZ} predict that $r_{G}=1$.

\
\subsection{A first look to the virial expansion}

The effective density of the particles is proportional to $r^D$ and excluded volume 
effects becomes larger when $r$ increases.  The usual virial expansion may be used to 
compute $S(r)$ as function of $r$ and of the fugacity in the gran-canonical ensemble:
\be 
{\cal Z}(V,z)= \sum_N {z^N \over N!} Z(V,N) \equiv \exp(-V F(z)) \ .
\ee

Eventually we can adjust the fugacity $z$ in such a way to have density one.  
These computations are well described in the literature. Diagrams are classified according to
increasing complexity. 

Our aim is to compare the  virial expansion for the partition function with the formulae 
that we have obtained for $\ln(P(0,r))$.
Let us consider firstly the results that we obtain if we consider a particular class
of diagrams, i.e. those of a chain.
In this case one finds that
\be
-S(r)= V(R) +\sum_{k=3,\infty}\int \prod_{i=1,k} d\nu(x_i)  \delta(\sum_{i=1,k} x_i) 
\ ,
\ee
where
\be
d\nu(x) = d^Dx \theta(R-x) \ .
\ee
We would like to compute the quantity $S_L(r)=\equiv \ln(P(0,r))$. We can follow the same arguments as in the section (\ref{SUM}). One finds, among many other terms
the following terns (neglecting factors 2)
\be
-S(r)=V(R) + \sum_{k=2,\infty}\int \prod_{i=1,k} \left( d\nu(x_i)\sum_{q_i}\right) \delta(\sum_{i=1,k} x_i 
q_i) \ ,
\ee
where the sum over the $q$ is done with the condition:
\begin{itemize}
\item
	All $q$ are different from zero.
\item
Conventionally $q_1$ is positive.
\item
The set of $k$ variables $q_i$ for $i=1,k$ is irreducible, i.e. they do not have common factors.
\item 
For $k=2$ the terms $q_1=1=\pm q_2$ are absent.
\end {itemize}

At each given $k$ the leading term in the sum comes from those term where $q_1=1$ and
$q_i=\pm1$. If we retain only these contributions we find (apart from factors two) exactly the same
result that for the pressure. 
Therefore we find that within the  
approximation of considering only the terms coming from one linear constraint and the for each 
moment taking only the leading term, the expression for $S_{L}(r)$ coincide with that 
for ${S(r)}$ obtained summing a subset of diagrams of the virial expansion.

This result should not be surprising and it can be simply understood: we have seen that the distribution of the points of a random lattice are random  if they do not satisfies 
linear constraints (see section (\ref{REM})) 
\footnote{This is clearly related that the matrix $\Lambda$ contains a very large number of parameters when the dimensions D goes to infinity and that a finite number of points may 
be points of the lattice}. This implies that as soon as we consider only the probability distribution of a finite (not diverging with $D$) number of points that would be just the 
same of points that are not constrained to be on the lattice.

\subsection{A naive conjectures}
Let us consider the generic contribution to the the virial expansion. 
 Every diagrams of the virial expansion of  $S(r)$ with $k$ lines and $L$ loops can be written 
be written in the form: 
\be
\int \prod_{i=1,k} d\nu(x_k) \prod_{l=1,L} \de^D\left(\sum_{i=1,k} M_{l,i}x_{i}\right),
\ee
where the quantities $M$ take the value 0 or 1. The form of the matrix $M$ fixes the 
topology of the diagram.

It is possible to check that all the contributions of the previous form also appears in 
the computation of $S_{L}(r)=-\ln(P(0,r))$, and the terms with many loops correspond to terms with 
more than one linear constraint.
All the diagrams of the virial expansion appear in the computation of $S_{L}(r)$, 
however this last expression contains much more terms that are not present in the virial expansion. 

It is possible that the leading contributions 
in both problems  are the same (as it happens in the case of the chain 
approximation) so one could may be tempted to conjecture that in the limit of infinite dimensions the two
problems coincide and in the infinite dimensional limit we have that
\be
S(r) \approx S_{L}(r) \ .
\ee

The previous equality would implies that in infinite dimensions the  maximal density of lattice packing is
equal to maximal density of packing without any lattice constraint. Although the conjecture may be true order by order in the virial expansion, we should be very careful because the 
sum of subdominant term may become dominant. I believe that this conjecture has only small chance to be true

\subsection{A more refined conjecture and open problems}

In order to further understand the problem we need to control the hard sphere gas in
large dimensions $D$ and eventually to treat the difference among the two problems as a
perturbation.  The task of computing the properties of the hard sphere gas in the infinite 
dimensional limit is non trivial (when
$r>1$) \cite{PS,PZ,TOR}.  In some sense we have transformed an unsolved problem into an other unsolved 
problem. The situation is not so bad, as far as the computation of the thermodynamical 
properties of an hard sphere gas may be not out of reach. One should firstly obtain some 
results in the liquid phase for $r>1$, and study the properties of the glass  transition is present.

We recall that some progresses have been recently done by resuming the contributions that correspond 
the hypernetted chain approximation in the infinite dimensional case for $r<e/2$ '\cite{PS}. Also the glass transition has been tentatively located around $r=1$ \cite{PZ}.

The situation is quite complex and it may be convenient to consider three functions for large dimensions $D$:
\begin{itemize} 
	\item $S(r))$ is the entropy for a gas of hard spheres as function of the radius. It is a function that may have singularities a various values of $r$. In three 
	dimensions it has a singularity at $r_{c}$ that corresponds to crystallization. It seem that in three dimensions at $r_{g}>r_{c}$
	 there is a glass transition, characterized by a faster increase of the pressure \footnote{As far as this phenomenon happens in the metastable phase, things are not so well
	 defined; however one can introduce slightly modified Hamiltonians that forbids crystallization \cite{PRUOCCO}.}.  It is possible than in higher dimensions the relative
	 position of the glass transition and the crystal transition do change.  It is also possible that in high dimensions the crystal like transition disappears, but this would be
	 quite unlikely. The value of $r_{c}$ is estimated to be equal to 1 \cite{PZ}.
	 \item $S_{L}(r)=\ln(P(0,r){D}$, .i. e.  the logarithm of the probability of finding a lattice packing with reduced radius $r$.  It should diverge toward infinity when we reach the
	 maximum density for a lattice packing, i.e. $r_{M}$.  The reader should notice that this function can be written as an integral over a finite dimensional space so for fixed
	 $D$ it should be a continuous function of the radius.  Only in the infinite dimensions it could develop a real first order phase transition, however it could be possible to
	 observe the premonitory signs of this transition by evaluating it numerically.  It is also not clear if the glass transition at $r=1$ is present in this contest.
	 \item The two previous quantities are not easy to compute, although they are well defined.  One could use here a third quantity that is maybe slightly less defined, but it
	 would be much easier to compute,i.e. $S_{E}(r)$ that is the approximate value of the entropy computed using integrals equations.  In other words $S_{E}(r)$ coincides with the 
	 results of the HNC approximation as soon the bridge diagrams are neglected.  An explicit solution of the HNC equation has been found in \cite{PS} in the region where $r<e/2$, 
	 i.e. where the Levenshtein bound is satisfied.
	 
	 If at some value of $r$ the bridge diagrams become important, one could consider  
	 more complex equation.  If only a finite number of bridge diagrams become relevant one after the other, this program could be implemented because if we have seen there
	 diagrams can be computed analytically in very large dimensions.
\end{itemize}

In this contest there are a few points to be investigated.
\begin{enumerate}
	\item
The computation of $S_{E}(r)$ is the more urgent task if we would like to get analytic predictions. While the situations is more or less clear in  the HNC, it is not clear if there 
is a value of the density where bridge diagrams must be considered.
\item One should use the analytic tools that are used to predict a glass transition in the hard sphere liquid, to see if they predicts also a phase transition in  $S_{L}(r)$.
\item One should compute the $S_{E}(r)$. in dimensions where the computation is feasible to see if there are sign of a first order transition.
\end{enumerate}
If we succeed in task 1, if the questions in two have a negative answer, one can put forward the conjecture that 
\be
S_{E}(r)=S_{L}(r) \ .
\ee
An important test of the conjecture would be to verify is the terms that we have neglected do not pile up.

In conclusion this paper has opened more problem that those it has solved, however  I am confident that strong progress can be done on this line. The difference and the 
commonalities of the two problems we are investigating could be further clarified.


\section*{Acknowledgment}
It is a pleasure for me to thank Jorge Kurchan and Marc M\`ezard for their contribution in the first steps of this work and for the many discussions on the  subject of this paper. 
I would also to thank Francesco Zamponi for many interesting discussion on this and related problems and Salvatore Torquato for interesting correspondence.

\section*{Appendix: on the large dimensional limit of some integrals}

In this approach it is crucial   to evaluate some integrals in
the limits where the dimension $D$ of the space goes to infinity. Most of the integrals we
consider go to zero exponentially.  This result is related to the fact that in large
dimensions most vectors are orthogonal. Here will mainly discussing the exponential factor, and we
will neglect the prefactors.

Before considering the general case let us start by studying a simple example:
\be
A=\int d\mu (x) d\mu (y)  \delta((x+y)^2-1),
\ee 
where
\be
d\mu (x) =S(D)^{-1} 2 \delta(x^2-1) d^Dx 
\ee
The factor 
\be
S(D)= {\pi^{D/2} \over \Gamma (D/2)}
\ee
is the surface of the $D$ dimensional sphere and it is such that
\be
\int d\mu (x)=1.
\ee

If we go to polar coordinates, we introduce the angle $\phi$ between$x$ and $y$ we find,
neglecting prefactor that are at most powers of $D$
\be
A \approx\int d \cos(\phi) \sin(\phi)^{D} \delta(\cos(\phi)- \frac12) \approx
(\frac34)^{D/2} \label{TRE}
\ee
The same result can be obtained if we  consider the same integral with the measure
\be
d\nu (x) =S(D)^{-1}   d^Dx \theta(1-x^2)
\ee
still satisfying the condition
\be
\int d\nu(x)=1.
\ee
We can also consider the integral
\be
A=\int d\nu (x) d\nu (y) S(D)^{-1} \theta((x+y)^2-1).
\ee 
This integral is proportional to the probability 
 of having three sphere with each sphere touching each of the remaining two.

In large dimensions the measure of a sphere is concentrated on its surface and the difference among
a $\theta$ or a $\delta$ function is irrelevant. Therefore the integral have a similar limit when the dimension goes to infinity.

The same result could be obtained by using the integral representation for the $\theta$ function
\be
\theta(z)= \int {d\al \over \al} \exp (i \al z),
\ee
where the integral is done from $-\infty$ to $\infty$, giving a small positive imaginary part to
$\al$.
In this way we obtain that
\bea
A \approx \Gamma(D/2)^{2} \int d\al_1 d\al_2 d\al_3 \exp (i\Sigma) (i\D)^{-D/2} = \\
 \Gamma(D/2)^{2} \int d\al_1 d\al_2 d\al_3 \exp (i\Sigma -D/2 \ln(i\D)) \ ,
\eea
where
\bea
\Sigma= \al_1+\al_2+\al_3,\ \ 
\D= \al_1\al_2+\al_2\al_3+\al_3\al_1 \ .
\eea
The saddle point equations are:
\be
i= { D ( \al_2+\al_3) \over 2 \D}\ ,
\ee
plus those obtained by permutations.

The solution to the saddle equations is given by
\begin{equation}
\al_1=\al_2=\al_3= i{ D \over 3}.
\end{equation}
Substituting back we find the result \form{TRE} .

Alternatively,  we can formally introduce the variables
\be
z_i={\al_i \over \Sigma} \for i=1,3
\ee
that satisfy the constraint
\be
\sum_{i=1,3} z_i=1.
\ee
We finally find 
\bea
A\approx 
\Gamma(D/2)^2 \int d\Si \Si^{-D-1} \exp (i\Si) dz_1 dz_2 dz_3 \delta(\sum_{i=1,3}z_i-1)\D^{-D/2}=\\
{\Gamma(D/2)^2 \over\Gamma(D)} \int  dz_1 dz_2 dz_3 \delta(\sum_{i=1,3}z_i-1)\D^{-D/2} 
\approx 2^{-D}\int  dz_1 dz_2 dz_3 \delta(\sum_{i=1,3} z_i-1)\D^{-D/2},
\eea
where
\be
\D= z_1 z_2+z_2 z_3+z_3 z_1
\ee
We finally find that the saddle point is given by
\be
z_i=\frac13,
\ee
and we recover the previous result.

The same technique can be  used to estimate the integrals
\be
A_N= \int \prod_{i=1,N-1} d\nu(x_i) \theta (1-(\sum_{i=1,N-1}x_i)^2) \label{AN}
\ee
We find 
\be
A_N \approx ({N \over N-1})^{N-1} ({1 \over N-1})^{D/2} \to_{N\to\infty} ({e \over N})^{D/2}.
\ee

Other integral may be estimated in the same way. We may be interested to consider the cluster
integral:
\be
C=\int d\nu(x_1)d\nu(x_2)d\nu(x_3) \theta(1-(x_1+x_2)^2)\theta(1-(x_2+x_3)^2)\theta(1-(x_3+x_1)^2).
\ee
This integral is proportional to the probability 
 of having four sphere with each sphere touching each of the remaining three.
We find that
\be
C\approx {\Gamma(D/2)^3 \over \Gamma(3D/2)}\int \prod_{i=1,6} 
dz_i\delta(\sum_{i=1,6}z_i-1) \D^{-D/2}
\ee
where
\bea
\D=
z_1 z_2 z_3 + z_1 z_3 z_4 + z_2 z_3 z_4 + z_1 z_2 z_5 + z_1 z_3 z_5 + z_1 z_4 z_5 + 
   z_2 z_4 z_5 + z_3 z_4 z_5 + \\
z_1 z_2 z_6 + z_2 z_3 z_6 + z_1 z_4 z_6 + z_2 z_4 z_6 + 
   z_3 z_4 z_6 + z_1 z_5 z_6 + z_2 z_5 z_6 + z_3 z_5 z_6
\eea
Finally using the saddle point 
\be
z_i= \frac16
\ee 
we find 
\be
C\approx \left(\frac12\right)^{D/2}.
\ee

 A related, but technically different problem arise in computing series of the form
  \be
  G \equiv \sum_{N=1,\infty} (-r)^{ND} A_N,
  \ee
 when $D$ is large and $A_{N}$ are defined in eq. \ref{AN}.  As soon as $r>1$ the series is divergent for large $N$.  On the other end the alternating sign structure implies a
 certain amount of cancellation. This kind of computation can be found in \cite{PS}, we report here for completeness.

  At this end it is useful to observe that using the convolution theorem, starting from the definition in that \form{AN} we find  that
  \be
  A_N =\int d^D p I(p)^N,
  \ee
  where $I(p)$ is the Fourier transform of the measure $d\nu(x)$.
  We thus find
  \be
  G= \int d^D p (1+r^D I(p))^{-1}\ .
  \ee

  After a short computation we find that $I(p)$ is proportional to a Bessel function. It is positive
  at small $p$ and it has a negative minimum at $p=p_m$, where $p_m$ behaves as
  \be
  p_m= A D+O(1), \ \ \ \ A=\frac12
  \ee
  The value of the function $I(p_m)$ can be estimated to be
  \be
  I(p_m)= -B^{-D}, \ \ \ B=\sqrt{e/2} \approx 1.166
  \ee

  This result for the function $I(p)$ can be obtained directly from the known formulae on Bessel
  functions. It may be instructive to derive it from scratch.

  We can start from the representation for $D$-dimensional Fourier transform in polar coordinates
  \be
  I(p)=\int_{-1}^1 dx (1-x^2)^{(D-1)/2} \exp ( i p x).
  \ee
  Alternatively using the same techniques as before we can write
  \be
  I(p)= {\pi^{D/2} \over V(D)}\int {d\la \over \la} \exp( -{p^2\over 4\la}) \la^{-D/2}
  \ee
  In both cases one finds a purely imaginary saddle point in the variable $x$ (or $\la$) that
  gives as usual a purely real contribution. For example the value of $x$ at the saddle
  point is given by
  \be
  x={2z \over 1+\sqrt{1-4z^2}}
  \ee
  where
  \be
  Dz =p
  \ee
  Therefore for positive $p<DA$ we finds that \be
  I(p)= \exp (-Dg(z)),
  \ee
  where the function $g$ is given by
  \be
  g(z)=\frac12\ln({1+\sqrt{1-4z^2}\over 2}) -{2 z^2 \over 1+\sqrt{1-4z^2}}.
  \ee
  On the contrary as soon as $p>DA$ the saddle point is no more purely imaginary and the result is
  complex. In the region where $p=DA+t$, $t$ being of order one when $D\to \infty$ one finds
  \be
  I(p)= \exp(-Dg(A) h(t)),
  \ee
  where $h(t)$ is an oscillating function of order 1.

 The chain approximation does not present difficulties for $r<B$, but it diverges as soon $r>B$ as the effect of the pile up of subsasyntotic terms.

%
%
	
 \end{document}